\documentclass[12pt]{amsart}

\usepackage[margin=1in]{geometry}

\usepackage[all,cmtip]{xy}
\usepackage{xspace}
\usepackage{amsmath}
\usepackage{amsthm}
\usepackage{thmtools}
\usepackage{amssymb}
\usepackage[framemethod=tikz]{mdframed}
\usepackage{url}
\usepackage{listings}
\usepackage{graphicx}
\usepackage{xcolor}

\def\rb{\rule[0.5mm]{1.2mm}{1.2mm}\xspace}
\graphicspath{{../}}

\theoremstyle{remark}

\newcommand{\bessie}{BESSIE}
\def\tool{\textsc{\bessie}\xspace}
\def\prob{\operatorname{prob}}
\def\coc{CoC\xspace}

\def\home{\texttt{home}\xspace}
\def\work{\texttt{work}\xspace}
\def\shopping{\texttt{shopping}\xspace}
\def\other{\texttt{other}\xspace}
\def\school{\texttt{school}\xspace}
\def\college{\texttt{college}\xspace}
\def\religion{\texttt{religion}\xspace}

\begin{document}
\title[\bessie]{\bessie: A Behavior and Epidemic Simulator for Use With Synthetic Populations}
\author{Henning S. Mortveit}
\address{
Department of Engineering Systems and Environment (SEAS) \and
Network Systems Science and Advanced Computing (BII), University of Virginia
}
\email[H.~S.~Mortveit]{henning.mortveit@virginia.edu}
\author[S.~Adams]{Stephen Adams}
\address{Intelligent Systems Division, Virginia Tech National Security Institute}
\email[S.~Adams]{scadams21@vt.edu}
\author[F.~Dadgostari]{Faraz Dadgostari}
\address{Network Systems Science and Advanced Computing (BII), University of Virginia}
\email[F.~Dadgostari]{fd4cd@virginia.edu}
\author[S.~Swarup]{Samarth Swarup}
\address{Network Systems Science and Advanced Computing (BII), University of Virginia}
\email[S.~Swarup]{ss7rs@virginia.edu}
\author[P.~Beling]{Peter Beling}
\address{
  Intelligent Systems Division, Virginia Tech National Security Institute  \and
  Department of Industrial and Systems Engineering, Virginia Tech
}
\email[P.~Beling]{beling@vt.edu}
\begin{abstract}
  In this paper, we present \tool (Behavior and Epidemic Simulator for
  Synthetic Information Environments), an open source, agent-based
  simulator for COVID-type epidemics. \tool uses a synthetic
  population where each person has demographic attributes, belong to a
  household, and has a base activity- and visit schedule covering
  seven days. The simulated disease spreads through contacts that
  arise from joint visits to the locations where activities take
  place.
  The simulation model has a plugin-type programmable behavioral
  model where, based on the dynamics and observables tracked by the
  simulator, agents decide on actions such as wearing a mask, engaging
  in social distancing, or refraining from certain activity types by
  staying at home instead.
  The plugins are supplied as Python code. To the best of our
  knowledge, \tool is a unique simulator supporting this feature set,
  and most certainly as open software.

  To illustrate the use of \tool, we provide a COVID-relevant example
  demonstrating some of its capabilities. The example uses a synthetic
  population for the City of Charlottesville, Virginia. Both this
  population and the Python plugin modules used in the example are
  made available.
  The Python implementation, which can run on anything from a laptop
  to a cluster, is made available under the Apache 2.0
  license\footnote{\url{https://www.apache.org/licenses/LICENSE-2.0.html}}. The
  example population accompanying this publication is made available
  under the CC BY 4.0
  license~\footnote{\url{https://creativecommons.org/licenses/by/4.0/}}.
\end{abstract}

\maketitle
\section{Introduction}
\label{sec:introduction}

Pandemics, such as COVID-19, are complex, behavior-driven phenomena~\cite{Singh2021}. Governments and other
authorities all over the world have relied on many non-pharmaceutical interventions (NPIs) for controlling
the spread of diseases, especially before vaccines became available. These interventions have included mandating
business and school closures (or reduced capacities), and promoting mask wearing, physical distancing, and
working from home~\cite{Desvars_Larrive_2020}. Broadly, the goal is to restrict interactions between people by
restricting mobility and to reduce the probability of transmission when people do interact.

In the agent-based modeling and simulation community, many models and platforms have been developed,  with
goals including understanding the spread of the epidemic, forecasting, guiding policy-making, and evaluating
counterfactuals~\cite{squazzoni2020computational, giabbanelli21covid}. While different goals may require different kinds of models,
doing a detailed analysis of the effects of NPIs requires two main ingredients or components: adequate detail
based on real data~\cite{swarup19adequacy} and true agency (adaptive coupling with the environment,
normativity, etc.~\cite{barandiaran09agency}).  This is because interactions in a population are contingent on, on the one hand, the demographics, activity
patterns, and built environment of a region, and on the other hand on human behavioral decisions,
which depend on descriptive and injunctive norms~\cite{Bicchieri_2020}, perceptions of risk and severity,
as well as perceptions of the efficacy of behaviors in risk-reduction~\cite{montano15behavior}.
In \cite{squazzoni2020computational} human behavioral modeling is also identified as one of the main challenges in agent-based modeling of the COVID-19 pandemic, as does the NSF in a recent Dear Colleague Letter~\cite{NSF_DCL_EPI:22}.

While many epidemic simulation models exist (see the Related Work Section), all the ones we know fail to satisfy at least one of the following aspects:
\begin{enumerate}
\item the model is agent-based;
\item it supports incorporating human behaviors through the notion of actions;
\item uses a detailed visit schedule that the agent may dynamically update;
\item the simulator and its source are made \emph{available} and
  \emph{accessible} for download under a suitable license;
\item has made input data \emph{available} and \emph{accessible} for at least one  non-trivial case under some suitable license;
\end{enumerate}
While there are many reasons why other research and development
efforts omit one or more of these elements, we remark that doing so
certainly presents real challenges for peer review, reproducibility,
and independent validation of models.

\textbf{Contributions.} Here we present the integrated epidemic/behavior simulator \tool. This computational tool integrates
(i) detailed synthetic populations, with
(ii) a completely configurable behavioral model governing the use of
personal protective equipment (PPE), social distancing, and options for abstaining from certain
activity types (e.g., non-essential activities), and
(iii) a COVID-like epidemic process where transmission may occur upon
contact arising when people simultaneously visit a common location.
Since agents in our model can base their behavior on observations of other agents and parts of their state (through the notion of local observables) as they visit locations, this can clearly give rise to ``clustered behaviors''. As far as we know, \textbf{this is a unique feature of \tool not present in any other simulator} as outlined in the Related Work section. We remark that \tool was developed to support a machine learning (ML) environment, and incorporating local observables in the actions decisions of agents was central to the model design.
The \tool simulator is implemented as a multi-process, shared
memory, Python package that can run on desktops and laptops, as well
as on clusters. Along with this tool, two synthetic population data
sets are provided: the hypothetical ``Smallville'' with its three
inhabitants that can be used for testing, and the City of
Charlottesville population which has about~41,000 people. Both
populations include a weekly visit schedule specifying what activities
the citizens conduct, when they do them, as well as where they take
place.

Our intention is to provide the scientific community with an open
simulator integrating a completely configurable human
behavioral model and an individual-based, location-based, extended
SEIR model whose parameters are configurable.
\tool can run on a desktop, laptop or a cluster taking
advantage of the number of available cores. We think this can provide
a model for publishing that supports both reproducibility and peer review.
\tool is available for download under the Apache~2.0 license through its git repository, see~\cite{BESSIE:21}. Its accompanying data, the synthetic population of the City of Charlottesville (\coc), is available for download under the CC-BY-4.0 license, see~\cite{COC:21}.

\textbf{Paper organization.} In the next section, we give
background and an overview of related work. This is followed by an
overview of \tool, including the synthetic populations, the epidemic
model, as well as the behavioral model. We then describe the usage at a high level,
illustrate its use through several application examples, and close with a discussion
of applications and possible extensions. The appendix sections contain a
detailed description of the design of \tool, including plugin
interfaces, input and output data formats, and the details of the
synthetic population structure.

\section{Related Work}
\label{sec:related}

There have been many simulations that have been developed and published since the early days of the COVID-19 epidemic,
including a large number of agent-based simulations. \cite{lorig2021} provide a recent review. They point out that, of the
126 models they review, only 6 include adaptive human behavior. While several include detailed synthetic populations, none
include both detailed synthetic populations and adaptive human behavioral models. Thus, \tool occupies a unique niche in
the space of agent-based simulators for COVID-19 (and for infectious disease epidemics in general). Our synthetic population is
also the most detailed of its kind, as it includes several demographic variables, household structure, data-derived weekly
activity schedules for all agents, and comprehensive activity location data~\cite{adiga15US}.
Here we briefly describe some of the more popular agent-based simulators that are comparable to our work, while also
highlighting the differences.

The IDM Covasim simulator~\cite{Kerr2020.05.10.20097469} is a widely used agent-based simulation of COVID-19. It uses
synthetic populations from \url{http://synthpops.org}, but works mainly on the agent interaction networks induced by the
synthetic population. These synthetic populations are also relatively simple, in that they do not model activity and mobility
patterns in the population or actual locations. They rely on data about age-stratified mixing in the population to generate
modeled contact networks for the simulation. This is, in fact, a micro-simulation model as opposed to an agent-based model,
since the ``agents'' in this simulation do not make observations, or do any decision-making. Many kinds of interventions
can be scripted, however, allowing simulation of many kinds of programmed scenarios.

The OpenABM-Covid19 simulator~\cite{hinch20openabm} is another open source COVID-19 simulator that has been widely used.
It is also Python-based, though the core components of the model are written in the C language for speed. It thus claims to
simulate a population of 1 million interacting people in seconds for each day of the simulation. It is designed primarily
to support policy-making in the UK, by simulating a default population of one million agents generated from demographic data
from the UK Census. In this case, the model of interaction in the population is somewhat stylized, as individuals move
between networks representing households, public transport, workplaces, schools, transient social gatherings, etc. These
locations do not appear to be based on data, though the average numbers of interactions are set to match age-stratified
mixing data. Once again, this is a micro-simulation, not a true agent-based simulation, capable of evaluating various
non-pharmaceutical interventions, including manual and digital contact tracing.

ComoKit~\cite{Gaudou2020} is also a well-known COVID-19 simulator based on the GAMA platform. It uses more sophisticated
synthetic populations than the previous two simulators, e.g., populations created by the Gen*~\cite{chapuis19genstar}
synthetic population generator. \cite{Gaudou2020} present a simulation of the Son Loi Commune in Vietnam, which has about
10,600 people and 3000 buildings. In their model, agents have week-long, hourly activity schedules, at various locations,
which is similar to our synthetic populations. They can also evaluate various policies such as social distancing, and the
simulator allows script-based editing of the types of agents, policies, and activities that are included in the simulation.
However, they also do not allow the individual agents to do their own decision-making, thus making this essentially a
micro-simulation.

While there are several other simulations (and simulators) that have been widely used (see~\cite{lorig2021} for many
examples), they largely share the characteristics of the ones described above. \tool is thus different from all of these
in allowing for decision-making by individual agents in a richly detailed environmental context. It thus opens the door to
connecting agent-based modeling efforts in computational epidemiology with research in behavior and decision-making from
domains such as psychology, sociology, and public health.

\section{\tool: Overview}
\label{sec:overview}
The \tool tool implements an agent-based, time-stepped model for
epidemic spread that integrates the following features:
\begin{itemize}
\item A \emph{synthetic population} with a visit schedule as detailed
  in the Synthetic Populations sub-section below.
  Such a population has a set of people
  with demographic attributes, and is partitioned into
  households. Each member of the population has an activity sequence
  where each activity is mapped to a location. By executing their
  activity sequence, people come into contact.
\item An \emph{epidemic process} taking place over the population, see
  the Epidemic Model sub-section.
  Transmission can occur when susceptible
  persons come in contact with infectious person(s) at one or more
  locations.
\item A customizable \emph{behavioral model} allowing each person to
  adopt a set of measures. These include wearing a mask, engage in social distancing, and modifying their visit schedule modifications to refrain from selected activities. Activities that are omitted are replaced by staying-at-home. The
  behavioral model can use demographic information about the person,
  as well as information about global observables (e.g., total number
  symptomatic people), or local observables capturing things they have
  seen at previous visits (e.g., number of symptomatic cases observed when shopping).
\end{itemize}
The following sub-sections provide details on the major \tool components.

\subsection{Synthetic Populations}
\label{sec:synthpop}

\tool uses the notion of a \emph{synthetic population} (SP) which is a
statistically accurate representation of a population of a given
region, see~\cite{Chen:20c,eubank:04} for a detailed
account. An SP for a given region~$R$ has the following components:
\begin{itemize}
\item The set of \emph{individuals}~$P$ of~$R$ grouped into
  \emph{households}. Each person is equipped with demographic
  attributes such as age, gender, household income, and a simplified
  version of the NAICS classification.
\item Each person has an \emph{activity sequence} specifying what
  activity they perform and when. The activity sequence covers a
  complete week and include the activity types covered are \home,
  \work, \shopping, \other, \school, \college and \religion.
\item A set of \emph{locations}~$L$ consisting of \emph{residence
  locations} and general \emph{activity locations}. Each household is
  assigned a residence location.
\item A location assignment~$\Lambda$ that for each person~$p$ and for
  each activity~$a$ of~$p$ assigns a location~$\ell \in L$.
\end{itemize}
As one can see, the synthetic population allows one to deduce
precisely who visits a location at the same time at any point
throughout the week. For \tool, this information is assimilated into a
\emph{person file} and a \emph{visit file}, see
the appendix section on usage
for a detailed description of the data records.

\subsection{Epidemic Model}
\label{sec:epimodel}

The epidemic model used in \tool is an extended $SEIR$ model that
includes the following states:
\begin{itemize}
\item $S$ -- susceptible;
\item $E$ -- exposed;
\item $I_s$ -- infectious and symptomatic;
\item $I_a$ -- infectious and asymptomatic;
\item $R$ -- recovered;
\end{itemize}
We write $\mathcal{X} = \{S, E, I_s, I_a, R\}$ for the set of health
states. The dynamics of the disease are split into two components: (i)
\emph{disease transmission} in which a susceptible person becomes
exposed when in contact with an infectious person, and (ii)
\emph{disease progression} covering all other health state changes. One may
regard transmission as the ``between people'' and progression as the
``within a person.''

\textbf{Disease transmission.} Transmissions may only occur when a susceptible person is in contact
with an infectious person. In general, we use the notion of a
\emph{transmission configuration} to capture the situation where a
susceptible person $p$ changes his/her state from $X_i$ (the entry
state) to~$X_j$ (the exit state) in contact with a person~$p'$ in an
infectious state~$X_k$ (the contact state), and denote this
configuration by the triple~$(X_i, X_j, X_k)$.
For our disease model there are only two transmission configurations:
  \begin{equation*}
    (S,E,I_s)\quad\text{and}\quad (S,E,I_a).
  \end{equation*}
For a general disease model, any entry state is called a
\emph{susceptible state}, any exit state an \emph{exposed state}, and
any contact state an \emph{infectious state}. Each transmission
configuration $(X_i, X_j, X_k)$ has an associated \emph{transmission
  weight} $\omega(X_i, X_j, X_k)$ that represents the relative weight
of this particular transition. It assumes the default value~1.0.
We remark that the transmission configurations are disease properties
independent of people.
Each state~$X_i \in \mathcal{X}$ has an inherent
infectivity~$\iota(X_i)$ and susceptibility~$\sigma(X_i)$. Note again that
these are person-independent disease parameters.
To capture individual factors (e.g., due to vaccinations and/or
wearing a mask), each person~$p$ is assigned an infectivity and
a susceptibility scaling factor, $\beta_\iota(p)$
and~$\beta_\sigma(p)$ respectively. These have a default value of~1.0 but are generally time-varying.
The \emph{effective susceptibility and infectivity} of a person~$p$ in
state~$X_i$ are modeled as,
\begin{align}
\sigma_p(X_i) &= \beta_\sigma(p) \times \sigma(X_i),\quad\text{and} \\
\iota_p(X_i) &= \beta_\iota(p) \times  \iota(X_i).
\end{align}
Finally, a disease model will have a \emph{transmissibility} that we
denote by~$\tau$. It may be regarded as a scaling factor. With this,
we can now define the \emph{propensity}~$\rho$ of a contact
configuration where persons~$p$ and~$p'$ are in contact at a
location~$\ell \in L$ for a duration~$T$:
\begin{equation}
\label{eq:propensity}
  \rho(p, p', (X_i, X_j, X_k), T, \ell) =
  \bigl[T  \tau\bigr] \times w_\ell \times
  \bigl[\beta_s(p)  \sigma(X_i)\bigr]  \times
  \bigl[\beta_i(p') \iota(X_k) \bigr] \times
  \omega(X_i, X_j, X_k).
\end{equation}

We use the Direct Gillespie Method \cite{Gillespie:76,Gillespie:77}
to determine if a disease transmission takes place when visiting a
location, and, if there are $m>1$ transmission configurations for a
susceptible person, which of the contact persons at that location to
whom one attributes the transmission. Specifically, for each person~$p$
in state~$X_i$ we set,
\begin{equation}
   A(P) = \sum_{p',j,k} \rho(p, p', (X_i, X_j, X_k), T, \ell),
\end{equation}
where the sum extends over all neighbors $p'$ of $p$ and indices $j$ and $k$
for which a transition may have taken place. Without loss of
generality, the index set of triples $K = \{(P', j, k)\}$ is well
ordered. To determine whether we have a transition we sample a random
number,
\begin{equation*}
   a = -\log(\operatorname{uniform}(0,1)) / A,
\end{equation*}
and if the inequality $a \le \mathcal{T}$ (the duration of the time
step) holds, we select the actual transition by sampling a uniform
random number $\alpha \in [0, A]$ and determine the index $\kappa \in
K$ for which,
\begin{equation}
  \sum_{\kappa - 1} \rho(p, p', (X_i, X_j, X_k), T, \ell)
    < \alpha
    \le \sum_{\kappa} \rho(p, p',
  (X_i, X_j, X_k), T, \ell) .
\end{equation}
Finally, for each time step, all such candidate transitions from all
locations are collected for each person~$p$, and the final infector
person~$p'$ is determined using uniform random sampling proportional
to the single propensity returned for each location. The state transition
for~$p$ will take place at the end of the each time step (and will
override any disease progression that may have been scheduled for the
current time step, see below.)
\medskip

\textbf{Disease Progression.} We use a \emph{disease progression
  diagram} to capture all possible health state transitions that take
place within a person in the absence of transmission processes and
interventions.
The diagram has nodes all possible health states~$\mathcal{X} =
\{X_i\}$ and directed edges~$e = (X_i, X_j)$ with assigned
probability~$p_e = \prob(X_i, X_j)$ and a \emph{dwell time
  distribution}~$D_e$. For each state $X_i$ it is required that
$\sum_j \prob(X_i,X_j) = 1$. The dwell time distribution~$D_e$ is the
probability density for the dwell time in health state~$X_i$ given
that the transition~$X_i \longrightarrow X_j$ will take place.

Algorithmically, the disease progression is determined upon entry to a
new state: as a person~$p$ enters a state~$X_i$, the next state $X_j$
is sampled according to the next state distribution induced by the
probabilities~$p_e$.
Once the next state $X_j$ is determined, the dwell time~$\Delta T$
(unit is simulation time steps) is computed by sampling from the
associated dwell time distribution~$D_{(X_i, X_j)}$, rounded to the
nearest integer if necessary, and bounded below by~0. If the current
time step is~$T$, the state transition is scheduled to take place at
the end of time step $T + \Delta T$. Note that in particular, this
means the transition can take place with the current time step.

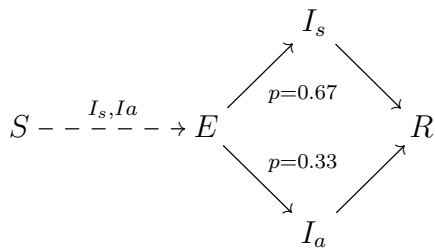
\begin{figure}[ht]
\centerline{
$\xymatrix{
                         &&                      & I_s \ar[rd]  &  \cr
S \ar@{-->}^{I_s, Ia}[rr] && E \ar[ru]_{p = 0.67}\ar[rd]^{p = 0.33} && R \cr
             &&                              & I_a \ar[ru] & }
$}
\caption{The combined transmission- and propagation diagram of the $SEIR$ model used by \tool. Dashed edges are for transmissions; solid edges are for propagation. The edges for transmission are decorated by the respective contact states.}
\label{fig:disease_diagram}
\end{figure}

\textbf{Example.} For the SEIR model of \tool, we have shown the combined transmission and progression diagram in Figure~\ref{fig:disease_diagram}. Note that transitions corresponding to transmissions are shown using dashed edges; edges corresponding to progression are shown as regular edges. Disease parameters can be found in the git repository in the file \hfill\break
\verb|./src/disease_model_v_1_0.py|.\footnote{Currently, disease parameters are not exposed in the configuration file, see the appendix sections.}

\subsection{Behavioral Model}
\label{sec:behavior}

The behavioral model is formulated using the notion of
\emph{actions}. Users may use the default model that is contained in
\verb|action_default_v_1_0.py|, or may supply their own model as
specified in Appendix~B.
Either way, at the beginning
of each time step, each person~$p$ will decide on what action(s) to
take in that time step for each of the following choices:
\begin{itemize}
\item Wear a mask ($a_1$);
\item Social-distance ($a_2$);
\item For each of the activity types other~($a_3$), college~($a_4$),
  shopping~($a_5$), religion~($a_6$), school~($a_7$), and work~($a_8$)
  decide whether or not each visit of that type for the given time
  step should be replaced by staying at home;
\end{itemize}
Thus, for each person $p$ and for each time step $t$, an action $a(p,t) = a = (a_1,
a_2, \ldots, a_8)$ is selected.
To decide on the actions to take, the following information is available for each person:
\begin{itemize}
\item the current time step~$t$;
\item their own current \emph{health state}~$s_p(t) \in \mathcal{S}$;
\item the static list of demographic variables of~$p$ (see the Usage section);
\item the list of \emph{global observables} containing the current
  total count $n(s)$ and fractions $r(s)$ of people in each of the
  disease states $s\in\mathcal{S}$ (e.g.,~$n(E)$ and~$r(E)$ for
  state~$E$);
\item the current \emph{local observables} for $p$. These capture the
  following information for each activity type $\alpha \in A$,
  recorded at the beginning of the most recent of visit for $p$ to a
  location to conduct an activity~$a$ of type~$\alpha$:
  \begin{itemize}
  \item the time step $t'$ when $a$ took place;
  \item the person ID (\verb|pid|), the location ID (\verb|lid|)
    $\ell\in L$, and the activity type $\alpha$;
  \item the total number of people present at $\ell$ at the start of
    that visit;
  \item the total count and fraction of people present at $\ell$ at
    the start of that visit of that are (i) symptomatic, (ii) are
    wearing a mask, and/or (iii) are engaging in social distancing.
  \end{itemize}
\end{itemize}
We remark that the actions of wearing a mask and social distancing
will be applied to all activities of the given time step. One may
argue that a person may decide that on an activity-by-activity basis; we may add
this in a future version of \tool.
Clearly, refraining from the activities of a given type will limit the
exposure in that context, but one may still be exposed to one's own
household members.
Wearing a mask or engaging in social distancing will cause the
person's infectivity and susceptibility scaling factors to be
reduced, currently as follows:
\begin{center}
\begin{tabular}{|l|l|l|l|}
\hline
Scaling factor & Value & Scaling factor & Value \\
\hline
\protect{\verb|mask_inf_scale|} & 0.8 & \verb|mask_susc_scale| & 0.8, \\
\verb|distancing_inf_scale| & 0.8 & \verb|distancing_susc_scale| & 0.8 \\
\hline
\end{tabular}
\end{center}

\section{Usage Overview}
\label{sec:usage-main}

The git repository contains a file \verb|Readme.md| with up-to-date
instructions for how to invoke \tool. For completeness, we include a
terse version here with section pointers to files and their precise
formats.
Running the \tool simulation tool requires having Python version~3.8
or greater installed on the system. Assuming the computing
environment is appropriately configured, an invocation of \tool will need the
following components, all of which are detailed in the appendix sections:
\begin{itemize}
\item a configuration file; 
\item a schema file;
\item a person file and a visit file; and
\item a Python action file containing a behavioral model.
\end{itemize}
The source distribution comes with a schema file, see the git
repository and the file \hfil\break
\verb|./schema/schema.json|. Using the provided example
of Smallville, which has a configuration file\hfill\break
\verb|./config/smallville_config_1_0_0.json|, one may invoke \tool
from the terminal as follows:

\qquad\verb|python -c smallville_config_1_0_0.json -s schema.json|

If the user is computing on a cluster using a job submission system (e.g.,
qsub or slurm), they may have to specify full paths for the
command-line arguments. \tool accepts an additional commandline
argument \verb|-l <level>| where \verb|level| is one of \verb|info|,
\verb|debug|, \verb|warning|, and \verb|error|. This may be useful if
there are problems running the code with a custom action file.

The person- and visit files are specified in the configuration file,
as is the Python file implementing the behavioral/action model. The
distribution comes with a basic behavioral model:

\qquad\verb|./src/action_default_v_1_0.py|

This may be a convenient starting point for constructing custom
versions. The user will most likely have to edit a copy of
\verb|smallville_config_1_0_0.json| to specify correct filenames and
paths for their system. We recommend using full paths in all cases. The
output will be generated in a directory specified in the
configuration file (\verb|output_directory|).

\section{Application Examples}
\label{sec:example}

We provide two example populations along with the \tool simulator:
(a) Smallville, and (b) City of Charlottesville (or CoC). Smallville is provided as an example that one should be able to run quickly on any machine. In this paper, we do not run analyses on Smallville, but we give hints of parameter values that may be interesting for exploration
and testing of action modules.
\subsection{Smallville}

In Smallville, everything is small-scale. On the other hand, its
citizens only work and stay at home, so presumably they get a lot
done. The person and visit files are constructed from the following
information:
\begin{itemize}
\item People: there are three citizens labeled 1, 2 and 3;
\item Locations: there are three residences with location ID
  (or \verb|lid|) of 11, 12 and 13. Additionally, there are three work
  locations whose \verb|lid|s are~1,~2 and~3;
\item Residence assignment: person \verb|i| lives in the residence
  location with \verb|lid = i+10|;
\item Activity sequences: each person has the same weekly activity
  sequence, and every day looks the same. Each person goes to work at
  midnight, work for one hour, and then returns to their home where they
  spend the remaining~23 hours of the day. People thus have only two
  activity types: \home and \work.
\item Activity location assignment. The only part that varies from day
  to day is the location where the citizens go for work:
  \begin{itemize}
  \item person 1: works at location 1 on Monday (day 0) through Sunday
    (day 6).
  \item person 2: works at location 1 on Monday through Wednesday. The
    remainder of the week, they go to location~2 for work.
  \item person 3: works at location~1 on Mondays, location~3 on
    Tuesdays and Wednesdays, and at location~2 on the remaining weekdays.
  \end{itemize}
\end{itemize}
Thus the most mixing that happens is at location~1. The weekly contact
network for the three citizens of Smallville is a complete
graph.\footnote{Despite the somewhat monotonous daily routines,
  Smallville consistently rates high in best-places-to-live reports.}
Parameters: for the verification and validation runs using Smallville, we used the model parameters~\verb|tau = 0.05| (transmissibility) and \verb|contact_probability = 1.0|. That way, simultaneous visits to the same location will imply that the corresponding people are in contact (or are connected by an edge in the induced contact network).

\subsection{City of Charlottesville (\coc)}

This example demonstrates the use of \tool using a synthetic
population for the City of Charlottesville, Virginia. The accompanying
action- and configuration files can be provided upon request
(see~\cite{BESSIE:21}) as can the person- and visit files~\cite{COC:21}.

\textbf{Model parameters.}
Here we use~\verb|tau = 0.0000015| and~\verb|contact_probability = 0.33|. These particular values were set to generate an unmitigated attack rate in the range 60--70\%, see~\cite{Buss:21}.
We remark that there are several \tool parameters in addition to transmissibility and contact probability that can be used for calibration, and there are also other metrics than attack rate that can be used on the dynamics,
examples including time-to-peak, and width-of-peak. Other calibration
parameters include:
\begin{itemize}
\item State infectivity and state susceptibility, \verb|iota| and \verb|sigma|: each health state $S$, $E$, $I_s$, $I_a$, and $R$ has associated a pair of such values. These disease parameters are set to either 0 or 1 in \tool, but could potentially be any positive number. Currently, these disease parameters are not exposed in the configuration file.
\item Dwell time distributions associated to the transitions $E\longrightarrow I_a$, $E\longrightarrow I_s$, $I_s\longrightarrow R$ and $I_a\longrightarrow R$. These are set in the file \verb|./src/disease_model_v_1_0.py| and are currently not exposed in the configuration file, but see the appendix section.
\item When wearing a mask, infectivity and susceptibility are scaled by \verb|mask_inf_scale| and \verb|mask_susc_scale| which are both set to \verb|0.8| (parameters are not exposed).
\item When doing social distancing, infectivity and susceptibility are scaled by \hfil\break
  \verb|distancing_inf_scale| and \verb|distancing_susc_scale| which are both set to \verb|0.8| (parameters are not exposed).
\end{itemize}
We consider four scenarios that demonstrate the capabilities of \tool,
and that may also give some insight into what it takes to construct
successful interventions from a public policy perspective. More
complex examples can easily be constructed using the ones provided
here as templates or starting points. The action and configuration files used for these examples are listed in Section~\ref{sec:example_files}.

\begin{figure}[ht]
  \centerline{\includegraphics[width=0.45\textwidth]{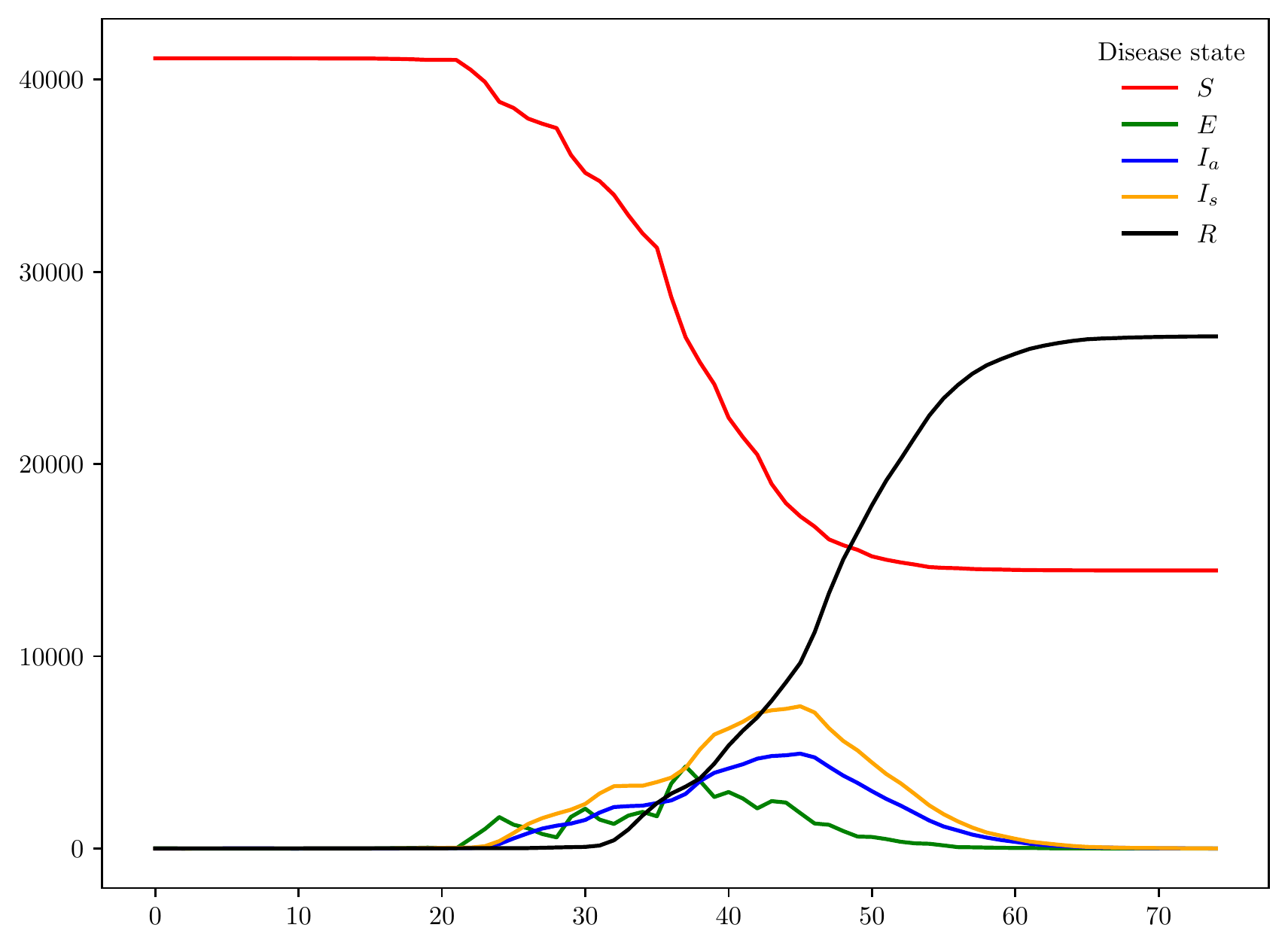}}
  \caption{Scenario 1 (base): in the base scenario, people go about their
    lives as if nothing happened. In this case, about 65\% of the
    population get COVID (black curve, recovered), comparable to values reported in~\cite{Buss:21}.}
  \label{fig:scenario1}
\end{figure}

\textbf{Scenario~1.} In this base scenario, everyone goes about their
visits as if nothing happened, not wearing masks nor doing any social
distancing.
As can be seen in the diagram of Figure~\ref{fig:scenario1}, the
end result for this simulation instance is that~$26,644$ people contract COVID during the course of this outbreak, corresponding to about~65\% of the
population of \coc.

\begin{figure}[ht]
  \centerline{
    \includegraphics[width=0.45\textwidth]{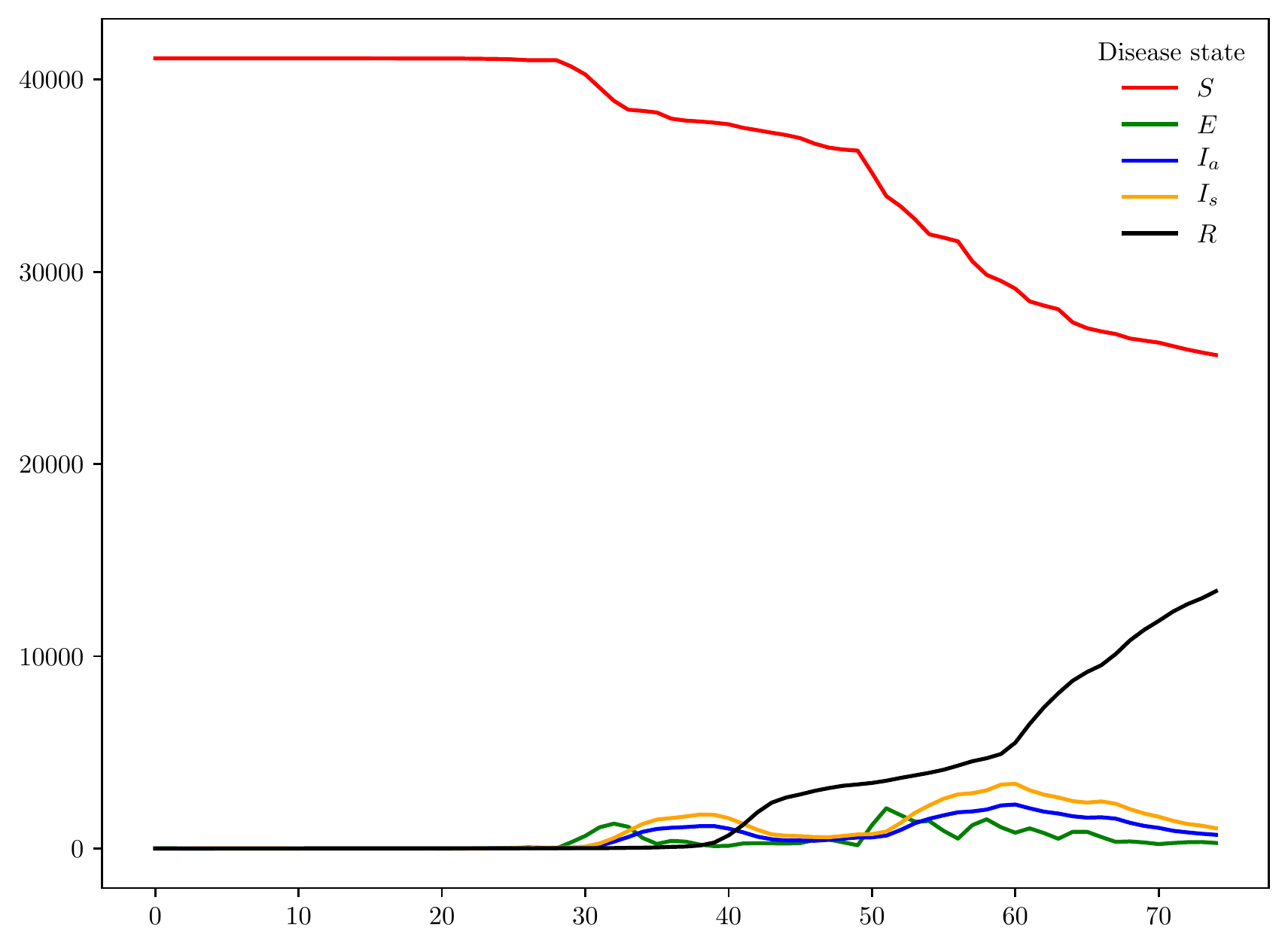}
    \quad
    \includegraphics[width=0.45\textwidth]{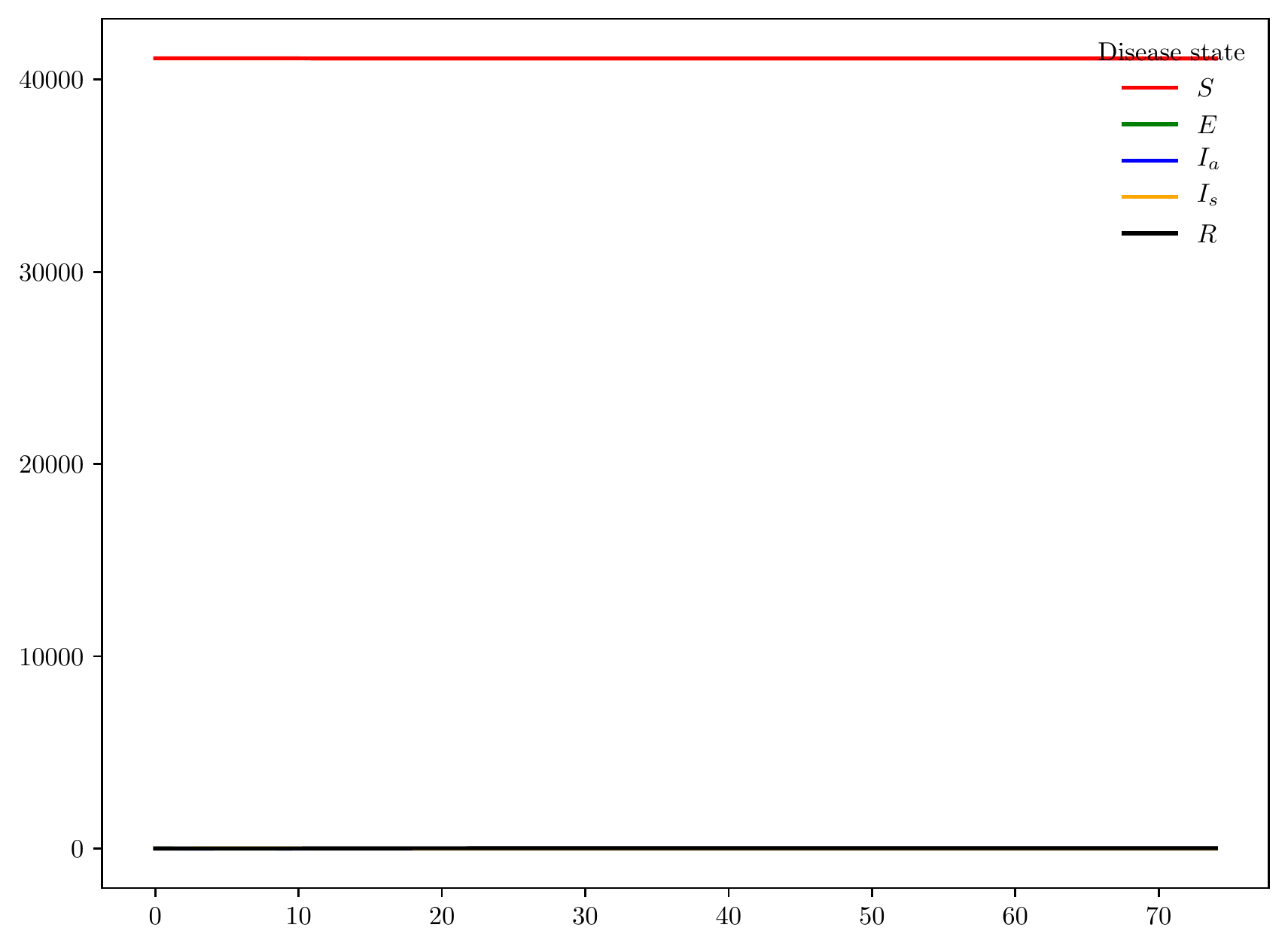}}
  \caption{Scenario 2: the diagram on the left shows the case where
    the subset is chosen at random at each iteration whereas the
    diagram on the right shows the case where a fixed random subset is
    chosen initially and used in every iteration. The difference is
    quite dramatic.}
  \label{fig:scenario2}
\end{figure}

\textbf{Scenario~2.} The second scenario demonstrates the case of wearing
masks and doing social distancing. In sub-scenario~(a), a subset
$P(t)$ of size 70\% of the population is chosen randomly at each
iteration $t$, and members of this set will wear a mask and do social distancing. In
sub-scenario~(b), a fixed but random subset $P$ of the population is
chosen. The members of $P$ will use a mask and do social distancing at
each iteration. Epidemic trajectories for each case are shown in
Figure~\ref{fig:scenario2}. We see that the epidemic is completely prevented for this particular simulation instance in sub-scenario~(b).

\begin{figure}[ht]
  \centerline{
    \includegraphics[width=0.45\textwidth]{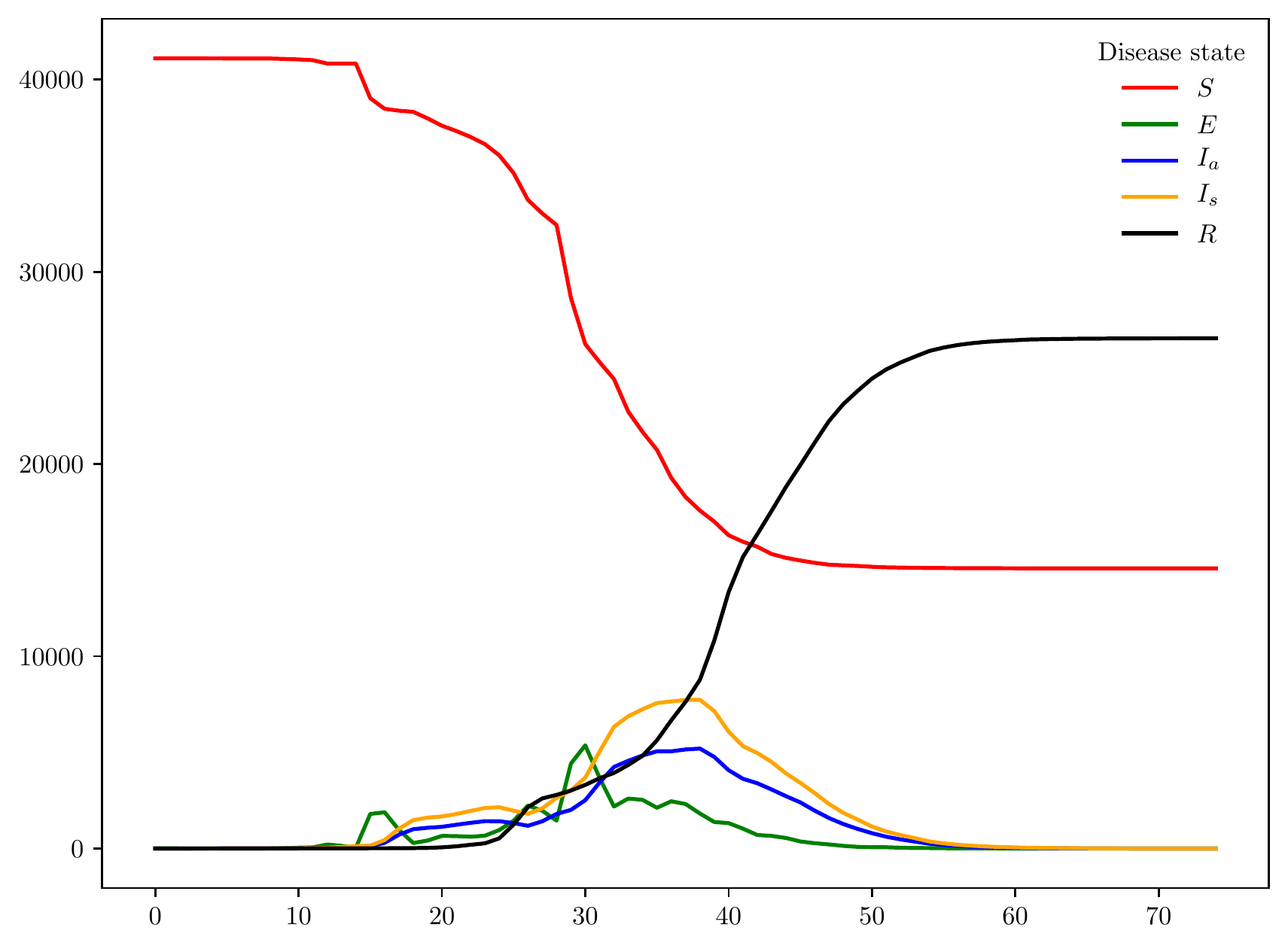}
    \quad
    \includegraphics[width=0.45\textwidth]{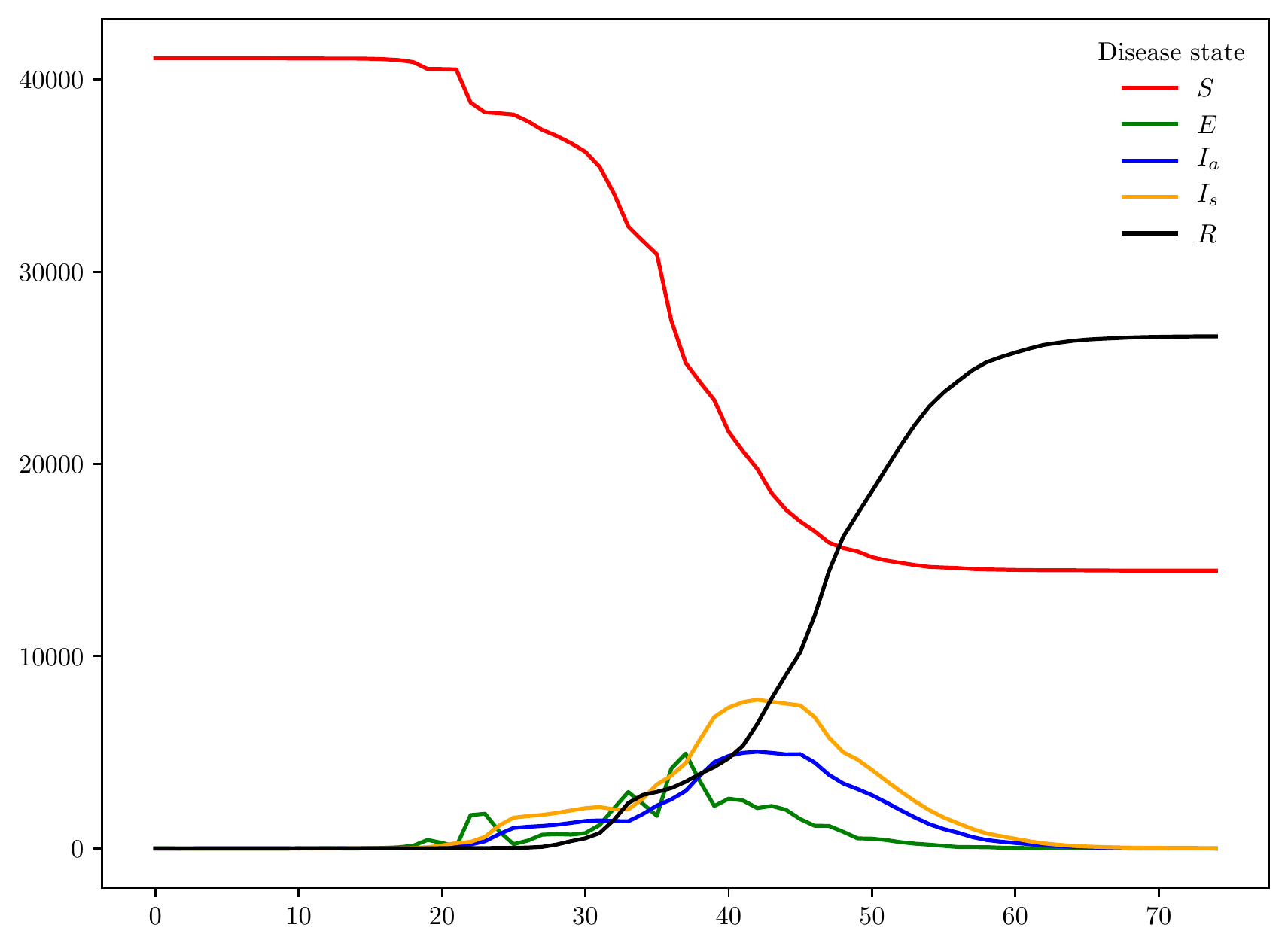}}
  \caption{Scenario 3: the left diagram shows the case where a random
    subset of size 75\% is chosen at every iteration starting on day 8
    and refrains from activities as explained, whereas the diagram on
    the right shows the case with a fixed random subset being used
    across all iterations. As can be seen, case (b) is virtually the
    same as case (a), except for a delay in the onsets.}
  \label{fig:scenario3}
\end{figure}

\textbf{Scenario~3.} In this case, we consider mandated drops of
activities. Starting at day~8, 75\% of people will re-route any visit
where the activity type is not \home or \work to their residence.  In addition, people with household income greater than or equal to~100,000 will conduct any work activity from their home. As for Scenario~2, we have two sub-scenarios where (a) the subset $P(t)$ is chosen randomly
at each iteration and (b) is chosen as a fixed subset $P$ that is used
in each iteration. In none of the cases do people wear masks or do
social distancing. The epidemic curves are shown in
Figure~\ref{fig:scenario3}. The main difference that emerges between these two sub-scenarios is that there is a delay in the onset of the epidemic in sub-scenario~(b).  This is not unexpected: since there is no coordination among household members, the disease can still propagate at people's homes.

\begin{figure}[ht]
  \centerline{\includegraphics[width=0.45\textwidth]{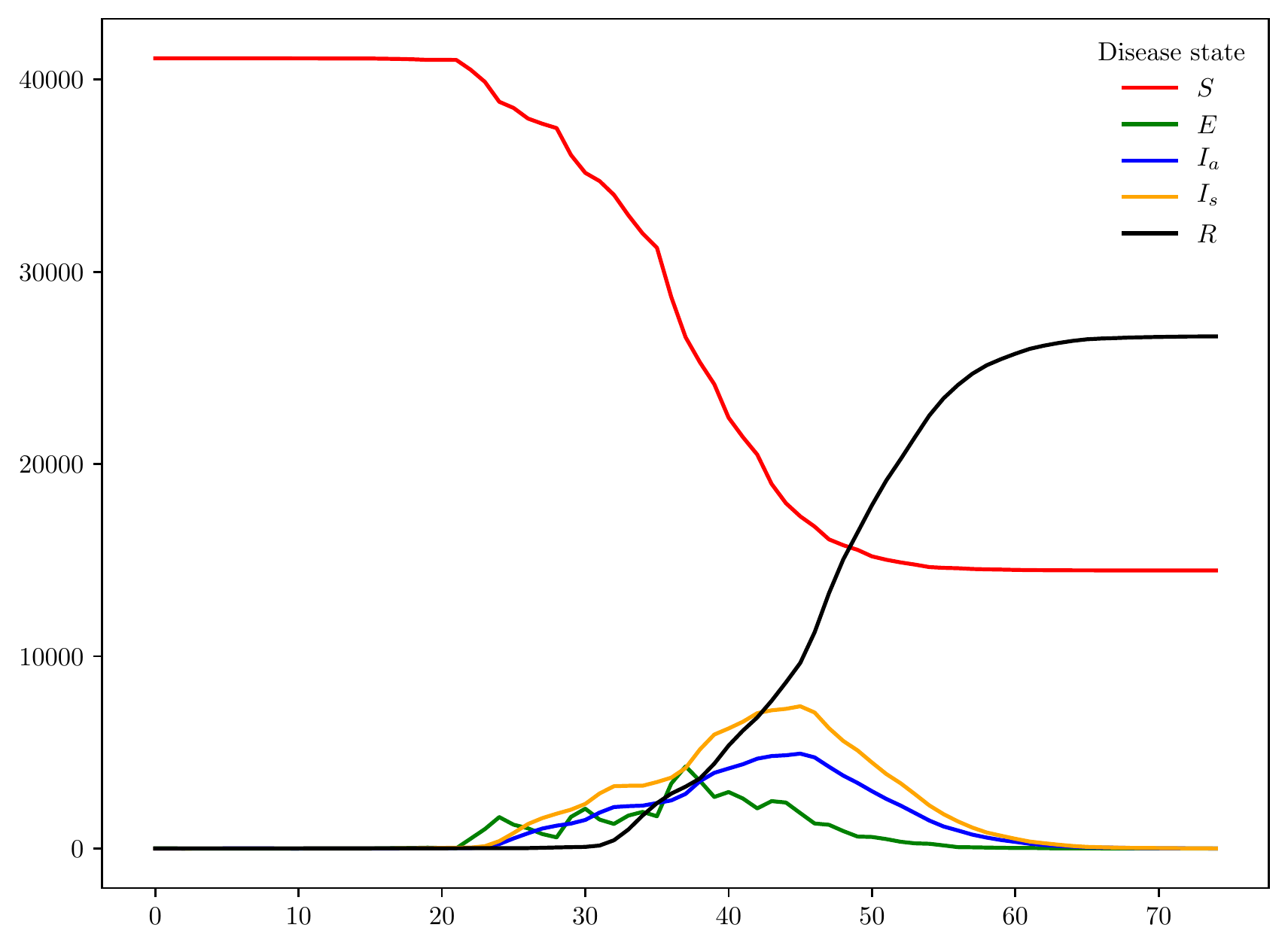}}
  \caption{Scenario 4 covers the case where people choose to remap certain visits based on their recent observables. The outcome is quite similar to the base case (Scenario~1).}
  \label{fig:scenario4}
\end{figure}

\textbf{Scenario~4.} This scenario demonstrates a case where actions are based on a person's local observables. For this, each person, at each iteration will consider their local observables for all non-\home activities. For \shopping, \other, \school, \college and \religion, if any of these observables is (1)
less than 7 days old and (2) contains at least 1 symptomatic case (including themselves), then the person will re-route any visit to such activity locations to their own residence. Additionally, a person whose household
income is at least~100,000 will conduct their work activity at their residence. For the latter, they also take into account the local observable for activity \work. Here the epidemic evolution is shown in Figure~\ref{fig:scenario4}.

\textbf{Summary.} While some of the observed dynamics may appear surprising at first glance, there are many causes. For example, in the case of Scenario~4, while dropping activities and staying at home may seem like it should have an impact, it does require coordination across the household: unless all members of the household consistently stay home, it is easy to see that even when one member does not observe symptomatic cases and do not modify their visits, then that can easily cause the entire household to fall sick. Moreover, if somebody records symptomatic cases in their observables, then very likely there are asymptomatic cases which are not detected. Basing one's actions on the cases one has observed in person may be a bad idea.

While the point of the scenarios is to demonstrate the capabilities of the \tool simulator (and these outcomes should certainly not be used to inform behavior or policy!), it does point to the fact that for interventions to be effective they require global planning and coordination.
Finally, we again want to emphasize that the epidemic parameters are not set by experts, and do not correspond to carefully calibrated values. If this was used to advise on policy formation then this step definitely would be included. The goal of showing these scenarios is to demonstrate some of the range and flexibility of the scenarios that can be handled by \tool.

\section{Discussion}
\label{sec:discussion}

In this paper we have introduced the \tool simulator and demonstrated some of its capabilities of several scenarios related to COVID and its intervention for a detailed synthetic population. As already alluded to in the Introduction, for a simulator to be able to address the example scenarios of Section~\ref{sec:example} at that level of detail, it would have to:
\begin{itemize}
\item use agents in the representation;
\item have an explicit representation of the locations where transmission occurs;
\item have a visit schedule for each person, and support dynamic changes to this visit schedule based on conditions on local observables, global observables, and a person's demographic attributes; and
\item track per-person local observables by activity across visits.
\end{itemize}
To the best of our knowledge, we are not aware of any tool that supports this, let alone offers up the code and the population data for download.

\subsection{Validation}
For the examples, we set contact probability and transmissibility to obtain an unmitigated attack rate near 60--70\%. While the focus of this work is to provide the \tool simulator, we remark that \tool has several parameters for the epidemics that can be used to calibrate to specific scenarios. In addition, the user can also introduce parameters in the action files they generate for the behavior models.

\subsection{Future directions}

\textbf{Speed and scaling.} \tool was implemented using Python. A main reason for this was to permit users to easily add action modules on their own through the \tool plugin interface. The complexity of this would be quite different with a language like C++. The architecture design for \tool is as a shared memory application running on multiple cores on a single compute node. This solution was used to work around the Global Interpreter Lock (GIL) in Python: a threaded approach would have been used in a language supporting concurrent threads.
The current solution also targets laptops and desktops which nowadays have at least two cores, and typically many more. If scaling and speed become a focus, a re-design and re-implementation in C++ using, e.g., MPI~\cite{MPI:2021} and/or OpenMP~\cite{OpenMP:2021} is a natural way to go. However, such a solution would effectively only cater to those who have access to a computing cluster or similar architectures.

\textbf{Intervention features.} As demonstrated in Application Examples section, it would be useful to be able to coordinate interventions efforts within a household, or more broadly, introduce actions for groups of agents based on their collective state. This, while certainly useful for epidemic simulations, would significantly change the scope of \tool which was to support learning for individual agents.
In related work, the authors have built
the EpiHiper C++/OpenMP/MPI \emph{network-based, epidemics simulator}~\cite{EpiHiper:21} has an intervention language that supports quite flexible construction of sets of people using predicates on their demographics and health state, and can then apply actions (i.e., state changes) to the members of these set. Quarantining or self-isolation of all members of selected households is therefore directly supported. If \tool were to be generalized in this direction, a similar approach will be used.

\textbf{Disease models.} Currently, the \tool disease model is limited to an extended \emph{SEIR} model. While flexible, there are certainly other or more refined models that can be considered. If this is needed in the future, it will be approached using a plugin design similar to that of the action/behavioral model that is already there.

\textbf{Calibration support.} Due to the scope of the design, several parameters are currently not exposed in the configuration file. Depending on future needs, a calibration framework can be added that and that (a) would expose relevant \tool parameters, and (b) would allow for efficient user declaration and exposure of custom parameters defined in their behavior models (and possibly custom disease models).

\subsection{Current limitations}

For a list of limitations and updates, we refer to the git repository and the files \verb|README.md| and \verb|./src/notes.txt|. The following are some known limitations that can cause issues if the user pushes boundaries and/or has complex computing environments.

\begin{itemize}
\item \tool requires Python version \verb|>=3.8| due to the use of the shared memory manager functionality.
\item The \tool simulator uses a fixed port number (i.e., 50000) for its shared memory manager. If multiple instances of \tool are run on the same compute node (not unlikely if you use a cluster), a port conflict will arise. In this case, the behavior of \tool is undefined. It can easily be avoided through suitable queuing instructions under job submission systems such as \verb|slurm| and \verb|qsub|. Similarly, there is a small risk that a different application is already running on one's computer and that also uses port~50000. This somewhat technical parameter may be exposed in the configuration file in a future version of \tool.
\item Since the location assignment is dynamic (it depends on people's actions), the size of the shared memory buffers used by the sub-processes to report observables back to the main process are not known ahead of time (see Figure~\ref{fig:bessie_design}). While it would be nice to auto-determine these buffer sizes, this is currently done manually and specified in the configuration file. The current values were set of \coc and were specified to permit even very skewed loads of people onto locations. If the user introduces other populations with visit schedules, they may have to adjust these memory buffer sizes accordingly. The buffer size for the local observables, for example, will equal the maximal number of observables reported back from each sub-process during any iteration of the simulation: this number will depend on (a) the locations assigned to each sub-process, and (b) the visitors to those locations, where the latter number generally depends on the actions taken by all the members of the population.
\end{itemize}

\subsection{DISCLAIMER}
\label{sec:disclaimer}

The \tool simulator, the code, and the synthetic populations are provided as is, and the user assumes all responsibility for any use thereof. \tool is provided under the Apache 2.0 license while the \coc synthetic populations is made available under CC-BY-4.0.

\section{Acknowledgments}
\label{sec:ack}

The authors thank their collaborators in Network Systems Science and Advanced Computing Division (BII) and in the Department of Engineering Systems and Environment for many discussions. In particular we thank Stefan Hoops for discussions and advice related to the software design and technologies used for \bessie.
This work was funded in part by the grant FI00026 ``Machine Learning Efficient Behavioral Interventions for Novel Epidemics'' awarded by the the Global Infectious Diseases Institute at the University of Virginia.

\section{Appendix A: \bessie~Design Details}
\label{sec:design}

In this section we describe details of the agent based model and the design of its implementation (i.e., the \tool simulator.) When preparing a detailed behavioral- or action model, knowing the precise order of the model components will be important.

\subsection{Organization of input data}

\begin{figure}[ht]
\framebox{\centerline{\includegraphics[width=0.8\textwidth]{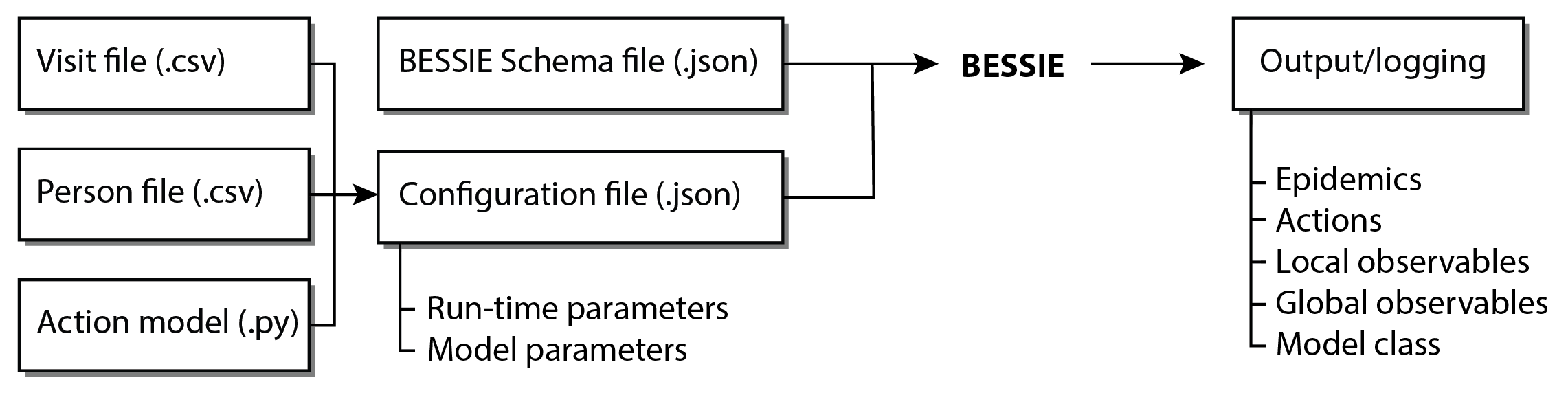}}}
\caption{An overview of the input data to \tool and their organization, including the configuration file and the configuration schema file.}
\label{fig:bessie_input}
\end{figure}

In Figure~\ref{fig:bessie_input}, the input data and its organization is shown. For the details of formats for the person file and the visit file, see the Person and Visit sub-section of Appendix B. The configuration file and the schema file are described in the Configuration File Format sub-section of Appendix B, while the requirements for the action file are given in the Action File sub-section of Appendix B.

\subsection{Architecture design}

\begin{figure}[ht]
\framebox{\centerline{\includegraphics[width=0.95\textwidth]{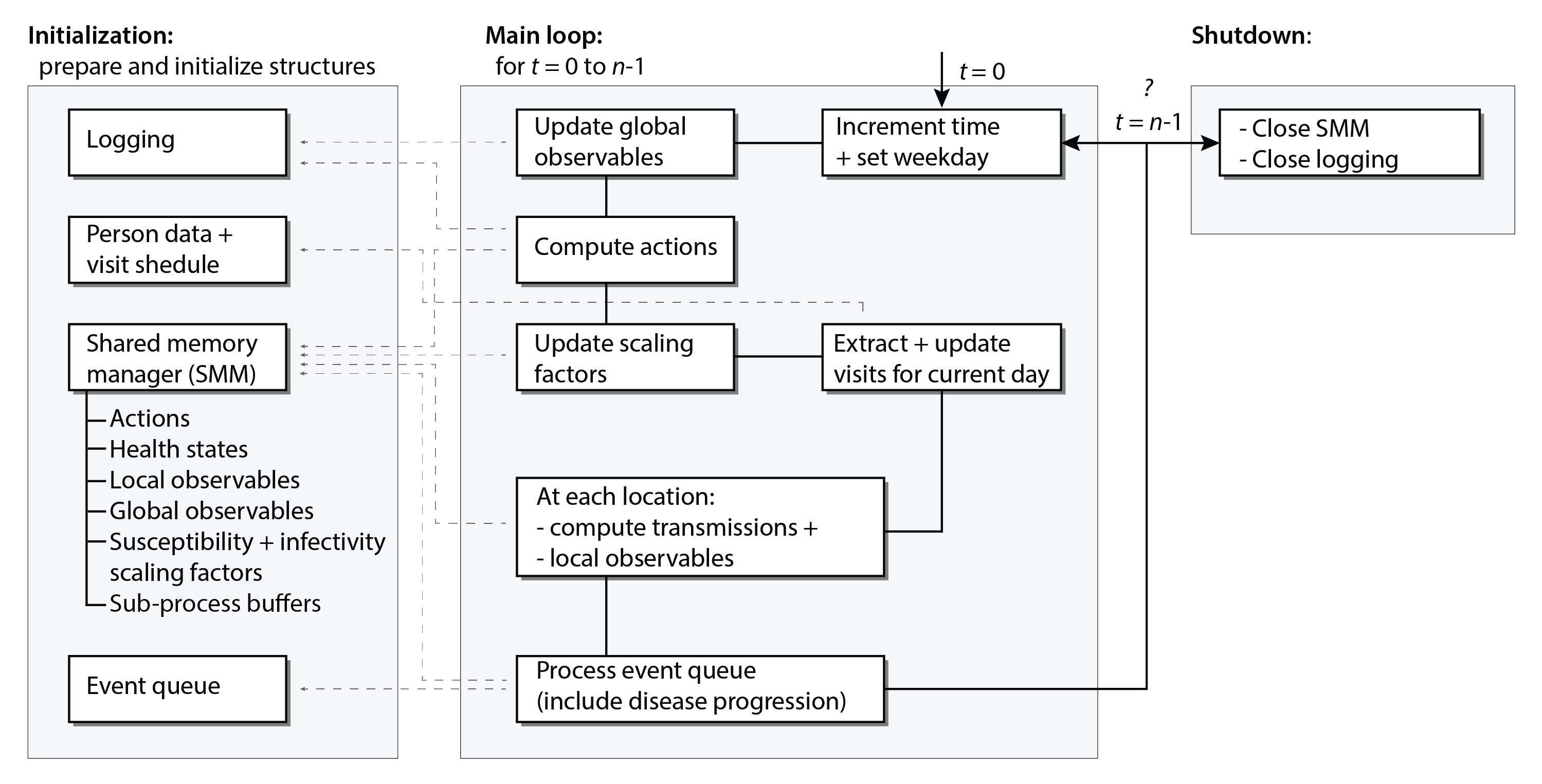}}}
\caption{An overview of the components in the design and the flow of computation for \tool. Dashed gray arrows indicate data dependencies as well as points where shared memory is being used as a data exchange point between the main \tool process and the sub-processes.}
\label{fig:bessie_design}
\end{figure}

A high level design of the simulator is given in Figure~\ref{fig:bessie_design}. Some notable points include the following:
\begin{itemize}
\item \textbf{Initialization:} Currently, one can only specify the number of people initially set to be in the exposed state~$E$. All others will have their state set to susceptible, or~$S$. This subset of the population is chosen at random. More elaborate methods for specifying the initial set of exposed people may be added later.
\item \textbf{Sub-process buffer sizes.} The shared memory manager has one buffer for each sub-process that is used for reporting back local observables. The minimal size of this buffer depends on the assignment of people to locations, and the partition of locations onto sub-processes. Since people may choose actions that re-map their visits to new locations, assessing a precise bound for this size is not straightforward. The current solution is to specify this through the configuration keys \verb|observable_max_items_per_iteration|. The value found in the example files for \coc are generously to make sure they work with~2 cores. A future version of \tool may include auto-estimation of this quantity. The same applies to the buffer that holds transmission events whose sizes are set by the key
  \hfil\break
  \verb|transmission_event_max_items_per_iteration|.
\item All arrays in shared memory use \verb|numpy| and user defined types as specified in \hfil\break\verb|./src/typedefs_v_1_0.py|. When developing custom action plugins, you will need to import this file. Examples are provided with the sources in the directory \verb|./models|.
\end{itemize}

\subsection{Notes about implementation}

The following is a list or simulator notes requirements.
\begin{itemize}
\item \tool requires Python version \verb|>=|~3.8 since it uses the \verb|SharedMemoryManager| component of \verb|multiprocessing|.
\item \tool has been tested on macOS Catalina, MS Windows, and on a Linux cluster environment, including under \verb|slurm|.
\end{itemize}

\section{Appendix B: Description of \tool\ Usage and Data Formats}
\label{sec:usage}

The input data and parameters to the \tool simulator are specified through a JSON configuration file. A JSON schema is used for validation, and is supplied with the source code of \tool. The syntax for invoking \tool is as follows:

\qquad\verb|python -c <config-filename> -s <schema-filename> [-l <info-level>]|
\medskip

Where \verb|info-level| is one of \verb|critical|, \verb|error|, \verb|warning|, \verb|info|, and \verb|debug|. The schema and configuration file formats are described next.

\subsection{Configuration File Format}
\label{sec:config}

\tool is designed to use a JSON configuration file with a matching
schema file as described in the Usage Overview section. In the case
of the supplied Smallville example, a configuration file may look like
the one below.

\textbf{Notes.}
\begin{itemize}
\item The \verb|num_procs| parameter should be set based on
your machine's specifics and its number of cores. For a laptop and
desktop, you may want to start modestly (e.g., using 2 cores) as fully
loading your cores can really tax your computer.
\item You will have to update the paths (\verb|base_dir| and
  \verb|pop_base_dir|) of the examples to match your computing
  environment.
\end{itemize}

\lstset{language=python} \lstset{tabsize=1}\lstset{basicstyle=\ttfamily\tiny}
\begin{mdframed}
\begin{lstlisting}
{
    "log_level" : "info",
    "platform": "rivanna",
    "version": "1_0_0",

    "config": {
        "run_parameters": {
            "executable": "bessie_v_1_0.py",
            "source_directory" : "{base_dir}/src",
            "logging_filename": "smallville.log",
            "output_directory": "./logs_smallville",

            "input" : {
                "person_filename" : "{pop_base_dir}/smallville/smallville_person_gidi.csv",
                "visit_filename" : "{pop_base_dir}/smallville/smallville_visits_gidi.csv"
            },

            "output" : {
                "action_logfile" : "log_action_logfile.txt",
                "modelclass_logfile" : "log_model_class.csv",
                "local_observable_logfile" : "log_local_observable_logfile.txt",
                "global_observable_logfile" : "log_global_observable_logfile.txt",
                "epidemics_logfile" : "log_epidemics_logfile.txt"
            },

            "observable_max_items_per_iteration": 1000,
            "transmission_event_max_items_per_iteration" : 1000,
            "num_procs" : 2
        },

        "model_parameters": {
            "random_seed": 1235,
            "num_initial_exposed" : 2,
            "num_days" : 5,
	    "tau" : 0.05,
	    "contact_probability" : 1.0,
            "user_defined_action_filename" : "jasss_example_base_v_1_0",
	    "user_defined_action_directory" : "{base_dir}/models/",
            "schedule_choice" : "week"
        }
    }
}
\end{lstlisting}
\end{mdframed}

\subsection{Schema File Format}
\label{sec:schema}

The JSON schema used with \tool can be found in the git repository as
\verb|./schema/schema.json|. Please use the git schema for the most
up-to-date version. We have omitted the listing since it is nearly two pages long.

\subsection{Action File}
\label{sec:action_file}

The action file is a \tool mechanism that permits the construction and addition of custom action/behavioral models without any changes to the \tool core. It functions as a plugin in a manner quite similar to, e.g., a MS Windows DLL except that is is written in Python. The requirements for a plugin is that must contain the following two functions with the specified signatures:
\begin{verbatim}
    CreateModelClassFile(filename, person_np_rep, person_class_np_rep)

    ActionSelection(action_np_rep_i,
                    time_step,
                    weekday,
                    person_np_rep_i,
                    state_np_rep_i,
                    visits_np_rep_i,
                    observables_np_rep,
                    global_observables_np_rep,
                    person_class_np_rep_i)
\end{verbatim}
Arguments to the two functions rely on type definitions contained in \verb|typedefs_v_1_0| which corresponds to the \tool file
\verb|./src/typedefs_v_1_0.py|. Thus you likely will want to add
\begin{verbatim}
    import typedefs_v_1_0 as typedefs
\end{verbatim}
near the beginning of your action action file. Your file may reference other files from your action file as long as they are present in the same directory as the action file (\tool will add the directory you specify for your action file to the search path -- recommended) or if you update your Python path accordingly.

\textbf{CreateModelClassFile.} This function provides the option to accomplish two things: (i) it can assign a \emph{person class} (integer) to each person based on their demographic attributes as specified in the person file (provided in the \verb|person_np_rep| argument), and (ii) it can write out this person class to a CSV file as specified in the Output Data section. This file is not used by
\tool, but may be useful for analyzing the outcome. The generated person class, however, will be passed to the \verb|ActionSelection| (see below). This aspect may be useful, since the plugin cannot (at least not easily) retain state across iterations. We refer to the examples for use cases.
\begin{itemize}
\item \verb|filename|: name of CSV file to construct (type:
  string)\hfil\break
\item \verb|person_np_rep|: a \verb|numpy| array with elements of type
  \verb|person_dtype| as defined in the repository file
  \verb|./src/typedefs_v_1_0.py|. The order is exactly the same as in
  the supplied person file.
\item \verb|person_class_np_rep|: a \verb|numpy| array with elements of type
  \verb|person_class_dtype| as defined in the repository file
  \verb|./src/typedefs_v_1_0.py|. The order is exactly the same as in
  the supplied person file.
\end{itemize}

\textbf{ActionSelection.} The action selection function is called at every iteration for each person in the course of the simulation. This provides the mechanism that one can use to specify which of the actions a person should (or should not) adopt at each iteration. Its argument are as follows:
\begin{itemize}
\item \verb|action_np_rep_i|: a single element of type
  \verb|action_dtype| as defined in \hfil\break\verb|./src/typedefs_v_1_0.py|. The function will set each element of this variable (they correspond to the list of possible actions). These values will be passed back to \tool.
\item \verb|time_step|: the current time step (or iteration) starting at \verb|t=0|.
\item \verb|weekday|: the current weekday. Here Monday corresponds to 0, Tuesday to $1$, and so on.
\item \verb|person_np_rep_i|: as single element of type
  \verb|person_dtype| holding the demographic attributes of the person as specified in the person file.
\item \verb|state_np_rep_i|: the health state of the person.
\item \verb|visits_np_rep|: the complete array of visits for all
  people and all days. Each entry is of type \verb|visit_dtype| as
  defined in \verb|./src/typedefs_v_1_0.py|
\item \verb|observables_np_rep_i|: the local observable array for
  person \verb|i|. To reference for example the current local
  observable for work, use\hfil\break
  \verb|observables_np_rep_i[ typedefs.obs_work_index ]|.
\item \verb|global_observables_np_rep|: the global observables
  array. To access for example the current number of symptomatic
  cases, use\hfil\break
  \verb|global_observables_np_rep[time_step]['obs_Is_abs']|
\item \verb|person_class_np_rep_i|: the person class of person \verb|i| as assigned in the\hfil\break \verb|CreateModelClassFile| function.
\end{itemize}
We recommend looking at the example file \verb|./src/action_default_v_1_0.py| from the \tool repository.

\subsection{Person and Visit Files}
\label{sec:population_data_format}

The representation of the synthetic population is captured by two
files, a \emph{person file} and a \emph{visit file}.

\textbf{Person file.} The person file contains demographic and
household information for each person, including details about their
residence.\footnote{Note that in the accompanying data for \coc,
  coordinates are all set to (0.0, 0.0) due to license terms that
  apply to the data that was used to construct this instance.} The person file
is a CSV file with the following header, shown with two example rows
of data\textbf{}.
\begin{mdframed}
  \begin{tiny}
\begin{verbatim}
hid,pid,age,sex,employment_status,race,hispanic,designation,hh_size,hh_income,\
workers_in_family,lid,longitude,latitude,admin1,admin2,admin3,admin4
2208253,5586585,38,1,4,1,1,military,6,55000,1,1001018209,-78.4884675,38.0430255,\
51,540,201,1
2208253,5586586,37,2,6,6,1,none,6,55000,1,1001018209,-78.4884675,38.0430255,\
51,540,201,1
\end{verbatim}
  \end{tiny}
\end{mdframed}

\textbf{Data dictionary.} Records with reference to PUMS have details
described in~\cite{PUMS:21}.\hfil\break
\rb\verb|hid|: the household ID of the person (type: integer)\hfil\break
\rb\verb|pid|: the person ID of the person (type: unsigned
integer)\hfil\break
\rb\verb|age|: the age of the person in years (type:
integer)\hfil\break
\rb\verb|sex|: the sex of the person (type: enumeration \{ 1: male, 2:
female\})\hfil\break
\rb\verb|employment_status|: employed or not, civilian or armed forces (PUMS:ESR)\hfill\break
\rb\verb|race|: race of householder (PUMS RAC1P variable)\hfill\break
\rb\verb|hispanic|: a Boolean (PUMS HISP variable) (type: Boolean \{0,1\})\hfill\break
\rb\verb|designation|: a person designation derived from the PUMS NAICSP variable (type: string)\hfil\break
\rb\verb|hh_size|: number of persons in family (PUMS NPF variable);\hfill\break
\rb\verb|hh_income|: household income in the past 12 months in local currency (PUMS HINCP variable);\hfill\break
\rb\verb|workers_in_family|: Workers in the family in the past 12 months (PUMS WIF variable)\hfill\break
\rb\verb|lid|: location ID of assigned residence (type: integer)\hfil\break
\rb\verb|longitude|: the longitude of the residence (type: float; see footnote)\hfill\break
\rb\verb|latitude|: the latitude of the residence (type: float; see footnote)\hfill\break
\rb\verb|admin1|: 2-digit US FIPS code for the state\hfill\break
\rb\verb|admin2|: 3-digit US FIPS code for the county\hfill\break
\rb\verb|admin3|: 6-digit US FIPS code for the census track\hfill\break
\rb\verb|admin4|: 1-digit US FIPS code for the block group\hfill\break

\textbf{Visit file.} The visit file captures the activity sequence and
the baseline locations of visit for each person. We say baseline since the locations of visit may change during the simulation based on the actions chosen.
In the case of \coc, the sequence spans a week, starting at midnight Sunday/Monday, a time point we will refer of as~$T_0$.  The visit file uses the CSV format with columns as follows, including two example rows:
\begin{mdframed}
\begin{verbatim}
daynum,pid,activity_number,activity_type,start_time,end_time,duration,lid
0,5586585,0,1,0,27900,27900,1001018209
0,5586585,2,2,28800,45900,17100,82246
0,5586585,4,4,46800,48000,1200,86726
\end{verbatim}
\end{mdframed}

\textbf{Data dictionary.}\hfil\break
\rb\verb|daynum|: the weekday on which the contact started
(type: enumeration \{0:~Monday, 1:~Tuesday, ..., 6:~Sunday\})\hfill\break
\rb\verb|pid|: \verb|pid| of person (see above)\hfill\break
\rb\verb|activity_number|: the activity number in the person's activity
sequence (type: integer)\hfil\break
\rb\verb|activity_type|: the activity conducted during the visit
(type: enumeration \{\verb|home|:~1, \verb|work|:~2,
\verb|shopping|:~3, \verb|other|:~4, \verb|school|: 5,
\verb|college|:~6, \verb|religion|:~7, \verb|transit|:~0\}; note that
all transit activities are omitted in the distributed data);\hfill\break
\rb\verb|start_time|: activity start time measured in seconds since~$T_0$ (type: integer)\hfill\break
\rb\verb|end_time|: activity end time measured in seconds since~$T_0$ (type:
integer)\hfill\break
\rb\verb|duration|: activity duration measured in seconds (type: integer)\hfill\break
\rb\verb|lid|: ID of location visited (type: integer)\hfill\break

\section{Output data}
\label{sec:output_data}
The \tool simulator generates five output files. These are all CSV files
with column headers and data dictionaries as described in the
following.

\subsection{Model class file} This is a classification of people into
model classes. The precise semantics of this will depend on the
behavioral model that you implement.\medskip
\begin{mdframed}
\begin{verbatim}
index,pid,model_class
0,5586585,1
1,5586586,1
\end{verbatim}
\end{mdframed}
\textbf{Data dictionary.}\hfil\break
\rb\verb|index|: the index used in the simulation for the \verb|numpy|
structures. It will equal the line number of the person's entry in the person file, not counting the header line (type: integer)\hfil\break
\rb\verb|pid|: the matching \verb|pid| field from the person file (see above)\hfil\break
\rb\verb|model_class|: the model class assigned to the person
under the given behavioral model specified in the action file (type: integer)\medskip

\subsection{Local observables file} This file contains one record for
each person for each non-\verb|home| activity type for each
iteration. In the case where there are multiple local observations for the same activity type within an iteration, this will contain the most recent local observable that was reported. This is a CSV file with the following structure:
\begin{mdframed}
  \begin{tiny}
\begin{verbatim}
iteration,obs_iteration,pid,lid,activity_type,n_total,symp_abs,symp_rel,mask_abs,\
mask_rel,distancing_abs,distancing_rel
0,0,5586585,0,1,0,0,0.0,0,0.0,0,0.0
0,0,5586585,0,2,0,0,0.0,0,0.0,0,0.0
0,0,5586585,0,3,0,0,0.0,0,0.0,0,0.0
\end{verbatim}
  \end{tiny}
\end{mdframed}

\textbf{Data dictionary.}\hfil\break
\rb\verb|iteration|: the iteration for the recorded observable (type: integer)\hfil\break
\rb\verb|obs_iteration|: the iteration for which the recorded
observable took place (type: integer)\hfil\break
\rb\verb|pid|: see above\hfil\break
\rb\verb|lid|: see above\hfil\break
\rb\verb|activity_type|: see above\hfil\break
\rb\verb|n_total|: total number of people present at the location,
including the person him/herself, at the start of their visit (type:
integer)\hfil\break
\rb\verb|symp_abs|: total number of people present at the location
whose health state was~\verb|Is|, including the person him/herself, at the
start of their visit (type: integer)\hfil\break
\rb\verb|symp_rel|: the ratio $\verb|symp_abs|/\verb|n_total|$ at the
start of the visit (type: float)\hfil\break
\rb\verb|mask_abs|: total number of people present at the location
wearing a mask, including the person him/herself, at the start of
their visit (type: integer)\hfil\break
\rb\verb|mask_rel|: the ratio $\verb|mask_abs|/\verb|n_total|$ at the
start of the visit (type: float)\hfil\break
\rb\verb|distancing_abs|: total number of people including the person
him/herself) present at the location that are social distancing
(keeping a distance of~6ft to others), at the start of their visit
(type: integer)\hfil\break
\rb\verb|distancing_rel|: the ratio $\verb|distancing_abs|/\verb|n_total|$, at the start of their visit (type float)\hfil\break

\subsection{Global observables file} The file contains the list of
global observables tracked at each iteration, represented as a CSV
file with the following structure:
\begin{mdframed}
  \begin{small}
\begin{verbatim}
iteration,S_abs,S_rel,E_abs,E_rel,Is_abs,Is_rel,Ia_abs,Ia_rel,R_abs,R_rel
0,41109,0.9997568130493164,10,0.0002431965694995597,0,0.0,0,0.0,\
0,0.0
1,41109,0.9997568130493164,8,0.0001945572585100308,0,0.0,\
2,4.86393146275077e-05,0,0.0
2,41074,0.998905599117279,36,0.000875507656019181,5,0.00012159828474977985,\
4,9.72786292550154e-05,0,0.0
\end{verbatim}
  \end{small}
\end{mdframed}
\textbf{Data dictionary.}\hfil\break
\rb\verb|iteration|: see above\hfil\break
\rb\verb|S_abs|: total number of people in health state \verb|S| (type: integer)\hfil\break
\rb\verb|S_rel|: fraction of of people in health state \verb|S| (type: float)\hfil\break
\rb\verb|E_abs|: total number of people in health state \verb|E| (type: integer))\hfil\break
\rb\verb|E_rel|: fraction of of people in health state \verb|E| (type: float)\hfil\break
\rb\verb|Is_abs|: total number of people in health state \verb|Is| (type: integer)\hfil\break
\rb\verb|Is_rel|: fraction of of people in health state \verb|Is| (type: float)\hfil\break
\rb\verb|Ia_abs|: total number of people in health state \verb|Ia| (type: integer)\hfil\break
\rb\verb|Ia_rel|: fraction of of people in health state \verb|Ia| (type: float)\hfil\break
\rb\verb|R_abs|: total number of people in health state \verb|R| (type: integer)\hfil\break
\rb\verb|R_rel|: fraction of of people in health state \verb|R| (type: float)\hfil\break

\subsection{Action file} The action log tracks the actions chosen by
each person at every iteration across the possible actions described in Section~\ref{sec:behavior}. It is a CSV file with the following
structure:
\begin{mdframed}
  \begin{small}
\begin{verbatim}
iteration,pid,mask,distancing,no_other,no_college,no_shopping,no_religion,\
no_school,no_work
0,5586585,1,1,1,1,0,1,1,1
0,5586586,1,1,1,1,0,1,1,1
0,5586587,1,1,0,0,1,1,0,1
\end{verbatim}
  \end{small}
\end{mdframed}
\textbf{Data dictionary.}\hfil\break
\rb\verb|iteration|: see above\hfil\break
\rb\verb|pid|: see above\hfil\break
\rb\verb|mask|: wear a mask at all locations visited during the
iteration (type: \{0,1\})\hfil\break
\rb\verb|distancing|: do social distancing (keep 6ft distance to all
others) at all locations visited during the iteration (type:
\{0,1\})\hfil\break
\rb\verb|no_other|: replace each activity of type \verb|other| by a
visit to their residence for the given iteration (type:
\{0,1\})\hfil\break
\rb\verb|no_college|: replace each activity of type \verb|college|
by a visit to their residence for the given iteration (type:
\{0,1\})\hfil\break
\rb\verb|no_shopping|: replace each activity of type \verb|shopping|
by a visit to their residence for the given iteration (type:
\{0,1\})\hfil\break
\rb\verb|no_religion|: replace each activity of type \verb|religion|
by a visit to their residence for the given iteration (type:
\{0,1\})\hfil\break
\rb\verb|no_school|: replace each activity of type \verb|school| by a
visit to their residence for the given iteration (type:
\{0,1\})\hfil\break
\rb\verb|no_work|: replace each activity of type \verb|work| by a
visit to their residence for the given iteration (type:
\{0,1\})\hfil\break

\subsection{Health state dynamics trajectory} This CSV log file contains a record
of each health state transition that took place for each person of the
population in the course of the simulation. Note that initialization is also considered as a state change event. As explained in Section~\ref{sec:epimodel}, there are \emph{transmission}- and \emph{progression} transitions. In the case of a transmission, the \verb|pid| of the infector (i.e., \verb|p2_pid|)
is included in the entry. For disease progression, we have always have~$\verb|p2_pid| = -1$. This CSV file has the following header:
\begin{mdframed}
\begin{verbatim}
iteration,state,p1_pid,p2_pid
-1,1,5586585,-1
-1,1,5586587,-1
-1,1,5586591,-1
-1,1,5586599,-1
\end{verbatim}
\end{mdframed}
\textbf{Data dictionary.}\hfil\break
\rb\verb|iteration|: see above; note that \verb|-1| means
initialization\hfil\break
\rb\verb|state|: the new health state (type: integer)\hfil\break
\rb\verb|p1_pid|: the \verb|pid| of the person whose health state
changed (type: same as \verb|pid|)\hfil\break
\rb\verb|p2_pid|: the \verb|pid| of the person deemed to have caused
the state transition in the case of a transmission (type: same as
\verb|pid|)\hfil\break

\subsection{Example action files and their configurations}
\label{sec:example_files}

The following is the list of files that were used as examples in this paper. We have listed both the action files (Python) and the corresponding configuration files (JSON). Note that you will have to adapt paths to fit your computing environment before trying the examples. Generally, you will find that the action file is the same across platforms; only the configuration file will need to be updated. In all the examples below, the configuration files are located in the directory \verb|./config/|, the the action files inside the directory \verb|./models/|. This part of the filenames have therefore been omitted in the listing of Table~\ref{tab:example_listing}.
\begin{table}[ht]
\qquad\begin{tabular}{|l|l|}
\hline
Example~1 (base) &
\verb|{ex}_base_v_1_0.py| \\
&
\verb|{ex}_base_config_rivanna_1_0_0.json| \\
\hline
Example~2a (mask/dist) &
\verb|{ex}_mask_distancing_v_1_0.py| \\
&
\verb|{ex}_mask_distancing_rivanna_1_0_0.json| \\
Example~2b  &
\verb|{ex}_mask_distancing_fixed_pid_set_v_1_0.py| \\
&
\verb|{ex}_mask_distancing_fixed_pid_set_rivanna_1_0_0.json|\\
\hline
Example 3a (visit drop) &
\verb|{ex}_visit_drop_mandated_v_1_0.py| \\
&
\verb|{ex}_visit_drop_mandated_rivanna_1_0_0.json| \\
Example~3b &
\verb|{ex}_visit_drop_mandated_fixed_pid_set_v_1_0.py| \\
&
\verb|{ex}_visit_drop_mandated_fixed_pid_set_rivanna_1_0_0.json| \\
\hline
Example 4 (observables) &
\verb|{ex}_visit_drop_observations_v_1_0.py| \\
&
\verb|{ex}_visit_drop_observations_rivanna_1_0_0.json| \\
\hline
Smallville (default) &
\verb|action_default_v_1_0.py| \\
&
\verb|smallville_config_rivanna_1_0_0.json| \\
\hline
\end{tabular}
\caption{The listing of example action files used in this paper along
  with matching configuration files. Here \texttt{\{ex\}} is
    short-hand for \texttt{jasss\_example}.}
\label{tab:example_listing}
\end{table}


\end{document}